\documentclass[conference,10pt]{IEEEtran}
\usepackage{epsfig,rotating,setspace,latexsym,amsmath,epsf,amssymb,amsfonts,bm,theorem,subfigure,epstopdf}
\usepackage{cite,authblk}
\usepackage{bbm}
\usepackage{algorithm}
\usepackage[noend]{algpseudocode}

\algrenewcommand\algorithmicforall{\textbf{foreach}}
\algrenewcommand\algorithmicindent{.8em}
\usepackage{color}
\usepackage{mathtools}

\newtheorem{lemma}{Lemma}

\newenvironment{Proof}[1]{\medskip\par\noindent{\bf Proof:\,}\,#1}{{\mbox{\,$\blacksquare$}\par}}
\IEEEoverridecommandlockouts
\allowdisplaybreaks

\begin{document}

\title{Partial Updates: Losing Information for Freshness  \thanks{This work was supported by NSF Grants CCF 17-13977 and ECCS 18-07348.}}

\author{Melih Bastopcu \qquad Sennur Ulukus\\
	\normalsize Department of Electrical and Computer Engineering\\
	\normalsize University of Maryland, College Park, MD 20742\\
	\normalsize  \emph{bastopcu@umd.edu} \qquad \emph{ulukus@umd.edu}}

\maketitle

\begin{abstract}	
We consider an information updating system where a source produces updates as requested by a transmitter. The transmitter further processes these updates in order to generate \emph{partial updates}, which have smaller information compared to the original updates, to be sent to a receiver. We study the problem of generating partial updates, and finding their corresponding real-valued codeword lengths, in order to minimize the average age experienced by the receiver, while maintaining a desired level of mutual information between the original and partial updates. This problem is NP hard. We relax the problem and develop an alternating minimization based iterative algorithm that generates a pmf for the partial updates, and the corresponding age-optimal real-valued codeword length for each update. We observe that there is a tradeoff between the attained average age and the mutual information between the original and partial updates.
\end{abstract}

\section{Introduction}
We consider a system where a transmitter updates a receiver while keeping the information at the receiver as fresh as possible. We use the metric of \textit{age of information} (AoI) to quantify the freshness of information. We define the instantaneous age at the receiver as the time elapsed since the most recently received update was generated at the transmitter. The AoI has been widely studied in queuing systems, energy harvesting and scheduling problems, multihop multicast networks, source and channel coding problems, and so on \cite{Kaul12a, Costa14, Bedewy16, He16a, Sun17a, Najm18b, Najm17, Soysal18, Soysal19, Yates17b, Hsu18b, Kadota18a, Gong19, Buyukates19c, Arafa19b, Sun17b, Sun18b, Bastopcu19, bastopcu20, Zou19b, Non_linear, Bastopcu18, bastopcu_soft_updates_journal, Arafa17b, Arafa17a, Wu18,Arafa_Age_Online, Arafa18f, Arafa19e, Farazi18, Yener_energy_19, Zhong16, Yates_Soljanin_source_coding, Mayekar18, partial_updates, Zhong17a, Zhong18b, Buyukates18, Buyukates18b, Buyukates19}.

\begin{figure}[t]
	\centerline{\includegraphics[width=0.95\columnwidth]{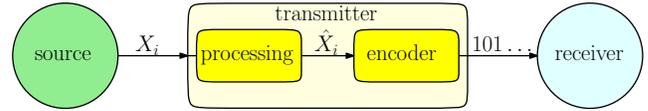}}
	\caption{An information updating system which consists of a source, a transmitter and a receiver.}
	\label{Fig:system_model}
	\vspace{-0.4cm}
\end{figure}

In our model, shown in Fig.~\ref{Fig:system_model}, a source generates updates as soon as requested by a transmitter. After an update is generated by the source, the transmitter further processes it to generate a partial update, and assigns a codeword to it using a binary alphabet. The transmitter sends this codeword to the receiver through a noiseless channel. Thus, the transmission time, i.e., the service time, for a partial update is equal to its codeword length. The average service time is equal to the expected codeword length, but the average age depends on both the first and second moments of the codeword lengths. Our goal is to optimize the partial update generation process, and the following codebook design, to minimize the age while maintaining a desired level of fidelity for the partial updates.

References that are most closely related to our work are \cite{Zhong16, Yates_Soljanin_source_coding, Mayekar18, MelihBatu1, MelihBatu3, partial_updates}. Reference \cite{Zhong16} considers age-optimal block code design and relates age to error exponents. Reference \cite{Yates_Soljanin_source_coding} considers the problem of assigning codewords to updates to minimize the average peak age. Reference \cite{Mayekar18} considers the problem of assigning real-valued codeword lengths to updates to minimize average age, and shows that Shannon codes can be used with a modified pmf to achieve asymptotically optimal performance. Reference \cite{MelihBatu1} considers the problem of \emph{selectively} encoding a given number of most probable updates while dropping the remaining least probable updates, and shows that this may yield better age performance than encoding all of the realizations. Reference \cite{MelihBatu3} considers the problem of sending an empty status update in the selective encoding scheme of \cite{MelihBatu1} to partially inform the receiver about the dropped updates, i.e., once an empty status update is received, the receiver knows that one of the dropped updates is realized but does not know which one specifically. Finally, reference \cite{partial_updates} introduces the concept of partial updates where both the information content and the transmission time are reduced compared to the original updates.

In this paper, we consider the problem of generating partial updates from the original updates, in a way to minimize the average age, while keeping the \emph{information content} of the partial updates at a desired level. In order to quantify the information similarity between the original updates $X$ and the partial updates $\hat{X}$, we use the mutual information between $X$ and $\hat{X}$. Since $I(X;\hat{X})=H(\hat{X})-H(\hat{X}|X)=H(\hat{X})$ as the partial updates $\hat{X}$ are functions of the original updates $X$, we need to minimize the age while keeping the entropy of the partial updates, $H(\hat{X})$, at a desired level. Thus, our problem reduces to finding a pmf for the partial updates $\hat{X}$ that can be generated from the given pmf of the original updates such that it yields the desired entropy, and the corresponding codeword lengths generated from the pmf of partial updates that minimize the age. This problem is NP-hard as we need to search over all partitioning of the realizations of the original updates to obtain the partial updates. We relax the problem to optimize over all pmfs for partial updates. While the resulting problem is non-convex, it is individually convex with respect to the pmf given the codewords lengths and vice versa. Thus, we develop an alternating minimization based iterative algorithm that optimizes one set of parameters (e.g., pmf) given the other set of parameters (e.g., codeword lengths). We investigate the tradeoff between the average age and the conveyed information content via numerical results.

\section{System Model and Problem Formulation}
We consider a communication system where a source generates independent and identically distributed status updates from a set $\mathcal{X} = \{x_1,x_2,\dots,x_n\}$ with a pmf $P_X(x_i) =\{p_1,p_2,\dots,p_n\}$ which is known. Without loss of generality, we assume that $p_i\geq p_j$ if $i<j$, i.e., the elements of the set $\mathcal{X}$ are sorted in decreasing order with respect to their probabilities. The transmitter requests an update from the source once the transmission of the previous update is completed. Thus, the source follows a generate at will model, and the transmitter follows a zero-wait model.

After an update is received by the transmitter, it further processes the update by using a function $g(X)$ to generate a partial update. The function $g(X)$ maps each update $x_i\in \mathcal{X}$ to the set of partial updates $\hat{\mathcal{X}}$, i.e., $g:\mathcal{X}\rightarrow \hat{\mathcal{X}}$, where the cardinality of $\hat{\mathcal{X}}$ is $k$, and $1\leq k\leq n$. When $k< n$, the transmitter maps some of the original updates from the set $\mathcal{X}$ to one partial update from the set $\hat{\mathcal{X}}$. When $k=n$, the transmitter sends the updates generated from the source without processing, i.e., $g(x)= x$ for all $x\in\mathcal{X}$. We write the pmf of  the partial updates as $P_{\hat{X}}(\hat{x}_i ) = \{\hat{p}_i |\hat{p}_i  = \sum_{i\in S_i}p_i, S_i = \{j| g(x_j) = \hat{x}_i, j=1,\dots, n\}, i= 1,\dots,k \}$.

For example, let us consider a source which generates an update from $\mathcal{X}= \{a,b,c,d\}$ with pmf $\{0.5,0.25,0.125,0.125\}$. The transmitter processes the updates to generate $k=3$ different partial updates. Let the set for the partial updates $\hat{\mathcal{X}}$ be $\{\{a\},\{b\},\{c,d\}\}$ with corresponding pmf $\{0.5, 0.25,0.25\}$. When update $a$ or $b$ is realized at the source, the receiver fully knows the realized update once the corresponding partial update is received. However, when update $c$ or $d$ is realized at the source, the partial update $\{c,d\}$ is transmitted. Once it is received, the receiver has the partial information about the update generated at the source, i.e., it knows that $c$ or $d$ is realized but does not know which one is realized specifically.

After generating the partial updates, the transmitter assigns codewords $c(\hat{x}_i)$ with lengths $\ell(\hat{x}_i)$ to each partial update $\hat{x}_i\in \hat{\mathcal{X}}$ by using a binary alphabet. Let the first and second moments of the codeword lengths be $\mathbb{E}[L]$ and $\mathbb{E}[L^2]$ where $\mathbb{E}[L] = \sum_{i=1}^{k}P_{\hat{X}}(\hat{x}_i)\ell(\hat{x}_i)$ and $\mathbb{E}[L^2] = \sum_{i=1}^{k}P_{\hat{X}}(\hat{x}_i)\ell(\hat{x}_i)^2$. We assume that the channel between the transmitter and the receiver is noiseless. Thus, if update $\hat{x}_i$ is transmitted, it takes $\ell(\hat{x}_i)$ units of time to deliver this partial update to the receiver, i.e., $\ell(\hat{x}_i)$ is the system service time for partial update $i$.

\begin{figure}[t]
	\centerline{\includegraphics[width=0.95\columnwidth]{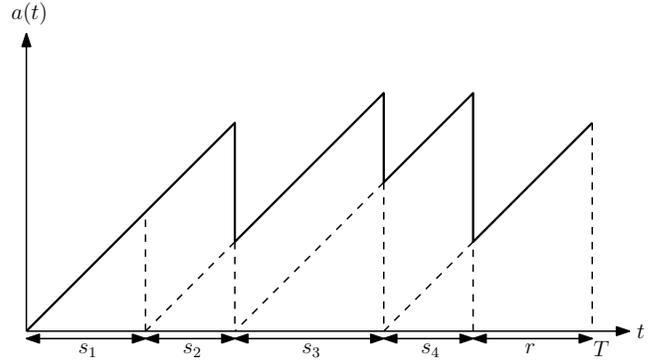}}
	\caption{Sample age evolution at the receiver.}
	\label{Fig:age_eval}
	\vspace{-0.4cm}
\end{figure}

In order to quantify the information retained by the partial updates, we use mutual information $I(X;\hat{X}) = H(\hat{X})-H(\hat{X}|X)$. We consider AoI to quantify the freshness of the information at the receiver. Let $a(t)$ be the instantaneous age at time $t$, with $a(0) = 0$. When there is no update, the age at the receiver increases linearly over time. When an update is received, the age at the receiver reduces to the age of the latest received update. Let $\Delta_T$ be the average AoI in the time interval $[0,T]$, which is given as
\begin{align}
\Delta_T =\frac{1}{T} \int_{0}^{T} a(t)dt,
\end{align}
and let $\Delta$ be the long term average AoI, i.e., $\Delta = \lim\limits_{T\rightarrow \infty} \Delta_T$. The age function evolves as in Fig.~\ref{Fig:age_eval}. Given that there are $m$ updates until time $T$, we write the average AoI, $\Delta_T$, as
\begin{align}
\Delta_T = \frac{1}{T}\left(\frac{1}{2}\sum_{i=1}^{m}s_i^2 +\sum_{i=1}^{m-1} s_i s_{i+1} + \frac{r^2}{2}+ \frac{s_N r}{2}\right),
\end{align}
where $r = T- \sum_{i=1}^{m}s_i$ and $s_i$ is the service time for the $i$th realized update. By using similar arguments as in \cite{Kaul12a}, we calculate the long term average AoI, $\Delta$, as
\begin{align}\label{age_expression}
\Delta = \lim\limits_{T\rightarrow \infty} \Delta_T = \frac{\mathbb{E}[S^2]}{2\mathbb{E}[S]}+\mathbb{E}[S],
\end{align}
where we use $\lim\limits_{T\rightarrow \infty}\frac{m}{T} = \frac{1}{\mathbb{E}[S]}$, $\lim\limits_{m\rightarrow \infty}\frac{\sum_{i=1}^{m} s_i^2}{2m} = \frac{\mathbb{E}[S^2]}{2}$ and $\lim\limits_{m\rightarrow \infty}\frac{\sum_{i=1}^{m-1} s_i s_{i+1}}{m} = \mathbb{E}[S]^2$. We note that the moments of the service times are equal to the moments of the codeword lengths, as service times in our paper are codeword lengths, and thus, we have $\mathbb{E}[S]= \mathbb{E}[L]$ and $\mathbb{E}[S^2] =\mathbb{E}[L^2]$.

In this paper, for a given $k$, our aim is to find the partial updates and the corresponding codeword lengths to minimize the average age while satisfying the constraints on the mutual information between the original and the partial updates, and the feasibility of the codeword lengths expressed by Kraft's inequality, i.e., $ \sum_{i=1}^{k} 2^{-\ell(\hat{x}_i)}\leq 1$. Therefore, we write the optimization problem as
\begin{align}
\label{problem1}
\min_{\{\hat{p}_i, \ell(\hat{x}_i) \}}  \quad &  \Delta \nonumber \\
\mbox{s.t.} \quad & I(X;\hat{X}) = \beta \nonumber \\
\quad & \sum_{i=1}^{k} 2^{-\ell(\hat{x}_i)}\leq 1 \nonumber \\
\quad & \ell(\hat{x}_i) \in \mathbb{Z}^+, \quad i\in\{1,\dots,k\}.
\end{align}
We study the optimization problem in (\ref{problem1}) in the next section.

\section{The Optimal Solution} \label{sect:opt_soln}
In this section, we solve a relaxed version of the problem in (\ref{problem1}) where we allow codeword lengths to be real-valued. We need to find the age-optimal pmf for the partial updates and the corresponding codeword lengths. We note that the constraint on the mutual information in (\ref{problem1}), $I(X;\hat{X}) =\beta$, is equivalent to $ H(\hat{X}) = \beta $ since $I(X;\hat{X}) = H(\hat{X})-H(\hat{X}|X) $ and $H(\hat{X}|X)=0$. Thus, using the age expression in (\ref{age_expression}), the relaxed optimization problem becomes,
\begin{align}
\label{problem3}
\min_{\{\hat{p}_i, \ell(\hat{x}_i) \}}  \quad &  \frac{\mathbb{E}[L^2]}{2\mathbb{E}[L]}+\mathbb{E}[L] \nonumber \\
\mbox{s.t.} \quad & H(\hat{X}) = \beta \nonumber \\
\quad & \sum_{i=1}^{k} 2^{-\ell(\hat{x}_i)}\leq 1 \nonumber \\
\quad & \ell(\hat{x}_i) \in \mathbb{R}^+, \quad i\in\{1,\dots,k\}.
\end{align}

Note that in order to solve the optimization problem in (\ref{problem3}), we need to find a partition of the original updates that produces the age-optimal pmf for the partial updates and the corresponding optimal real-valued codeword lengths. For a given pmf, $\hat{p}_i$, we obtain the age-optimal real-valued codeword lengths in Section \ref{subsect:code_design}.\footnote{We note that finding the age-optimal codeword lengths has been considered in \cite{Mayekar18}. Even though this is not our main contribution, we solve this problem here for completeness, and to present an alternative technique to \cite{Mayekar18}.} However, finding the optimal partition of the original updates is similar to a bin packing problem which is a well-known combinatorial NP-hard problem \cite{bin_packing}. Thus, the problem in (\ref{problem3}) is NP-hard and the optimal solution can be found by searching over all possible partitions.

In order to progress on the problem analytically, we relax the pmf constraint, and allow all possible pmfs for the partial updates. Note that originally the pmfs are limited only to the pmfs that can be generated from the partitions of $n$ original updates to $k$ partial updates; here, we allow all valid pmfs. Thus, we write the further relaxed problem as
\begin{align}
\label{problem4}
\min_{\{\hat{p}_i, \ell(\hat{x}_i) \}}  \quad &  \frac{\mathbb{E}[L^2]}{2\mathbb{E}[L]}+\mathbb{E}[L] \nonumber \\
\mbox{s.t.} \quad & H(\hat{X}) = \beta \nonumber \\
\quad & \sum_{i=1}^{k} 2^{-\ell(\hat{x}_i)}\leq 1 \nonumber \\
\quad & \sum_{i=1}^{k} \hat{p}_i= 1 \nonumber \\
\quad & \hat{p}_i\geq 0, \quad\ell(\hat{x}_i) \in \mathbb{R}^+, \quad i\in\{1,\dots,k\}.
\end{align}
Next, we define $p(\lambda)$ for the problem in (\ref{problem4}) as
\begin{align}
\label{problem4_mod}
p(\lambda) \vcentcolon \min_{\{\hat{p}_i, \ell(\hat{x}_i) \}}\quad &  \frac{\mathbb{E}[L^2]}{2}+\mathbb{E}[L]^2-\lambda \mathbb{E}[L] \nonumber \\
\mbox{s.t.} \quad & H(\hat{X}) = \beta \nonumber \\
\quad & \sum_{i=1}^{k} 2^{-\ell(\hat{x}_i)}\leq 1 \nonumber \\
\quad & \sum_{i=1}^{k} \hat{p}_i= 1 \nonumber \\
\quad & \hat{p}_i\geq 0, \quad\ell(\hat{x}_i) \in \mathbb{R}^+, \quad i\in\{1,\dots,k\}.
\end{align}
This approach was introduced in \cite{frac_programming} and has been used in \cite{Sun17b, Arafa_Age_Online}. One can show that $p(\lambda)$ is a decreasing function of $\lambda$ and the optimal solution is obtained when $p(\lambda) = 0$. The optimal age for the problem in (\ref{problem4_mod}) is equal to $\lambda$, i.e., $\Delta^* = \lambda$.

We introduce the Lagrangian function \cite{Boyd04} for (\ref{problem4_mod}) as
\begin{align}
\mathcal{L} = & \frac{\mathbb{E}[L^2]}{2}+\mathbb{E}[L]^2-\lambda \mathbb{E}[L] + \theta \left(\sum_{i=1}^{k} 2^{-\ell(\hat{x}_i)}- 1\right)\nonumber \\
&+\gamma\left( \sum_{i=1}^{k}\hat{p}_i \log \hat{p}_i+\beta\right)+\sigma\left(\sum_{i=1}^{k} \hat{p}_i -1\right)\nonumber \\
&- \sum_{i=1}^{k} \nu_i \hat{p}_i  - \sum_{i=1}^{k} \mu_i \ell(\hat{x}_i).
\end{align}
Next, we write the KKT conditions as
\begin{align}\label{KKT_cond2}
\frac{\partial\mathcal{L}}{\partial \ell(\hat{x}_i)} =& \hat{p}_i\ell(\hat{x}_i)+2\hat{p}_i\mathbb{E}[L] -\lambda p_i\nonumber \\ & -\theta (\log2)2^{-\ell(\hat{x}_i)} -\mu_i = 0,
\end{align}
\begin{align}\label{KKT_cond3}
\frac{\partial\mathcal{L}}{\partial \hat{p}_i} =& \frac{1}{2}\ell(\hat{x}_i)^2+2\ell(\hat{x}_i)\mathbb{E}[L] -\lambda \ell(\hat{x}_i)
\nonumber \\ &+\gamma\left( \log \hat{p}_i+\frac{1}{\log 2}\right)+\sigma -\nu_i = 0,
\end{align}
 for all $i$. The complementary slackness conditions are
\begin{align}
\theta\left( \sum_{i=1}^{k}2^{-\ell(\hat{x}_i)}-1\right) &= 0,\label{CS_cond2} \\
\gamma\left( \sum_{i=1}^{k}\hat{p}_i \log \hat{p}_i+\beta\right) & = 0, \label{CS_cond3}\\
\sigma\left(\sum_{i=1}^{k} \hat{p}_i -1\right) & = 0 ,\label{CS_cond4}\\
\nu_i \hat{p}_i& = 0, \label{CS_cond5}\\
 \mu_i \ell(\hat{x}_i)& = 0.\label{CS_cond6}
\end{align}

For the partial updates with $\hat{p}_i = 0$, we assign $ \ell(\hat{x}_i) = \infty$ as these updates never realize and we reserve the Kraft inequality for the updates with strictly positive probabilities. For the partial updates with $\hat{p}_i >0$, we have $ \nu_i = 0$ from (\ref{CS_cond5}). As entropy constraint $\beta>0$, we need at least two updates with $\hat{p}_i>0$. For each $\hat{p}_i>0$, we have $\ell(\hat{x}_i) >0 $. Otherwise, if we have $\ell(\hat{x}_i) =0 $ for an update with $\hat{p}_i>0$, other codeword lengths with non-zero probability have infinite lengths due to the Kraft inequality which clearly cannot be the optimal solution. Thus, we have $\ell(\hat{x}_i) >0 $ which implies $ \mu_i = 0$.

We note that the optimization problem in (\ref{problem4_mod}) is not convex as $\hat{p}_i$s and $\ell(\hat{x}_i)$s appear as multiplicative terms. However, for a given proper $\hat{p}_i$, the problem in (\ref{problem4_mod}) is convex with respect to the codeword lengths $\ell(\hat{x}_i)$. Similarly, for a given $\ell(\hat{x}_i)$, the problem in (\ref{problem4_mod}) is convex with respect to $\hat{p}_i$. We apply the alternating minimization method (see e.g., \cite{bertsekas, AlterMin, iterminimization2, Niesen07}) to find $(\hat{p}_i, \ell(\hat{x}_i))$ such that (\ref{KKT_cond2}) and (\ref{KKT_cond3}) are satisfied for all $i$. Starting with an initial pmf $\hat{p}_i$, we find the optimum real-valued codeword lengths $\ell(\hat{x}_i)$ for the initial pmf $\hat{p}_i$. Then, for given codeword lengths $\ell(\hat{x}_i)$, we find the pmf that is proper, i.e., $\sum_{i=1}^{k} \hat{p}_i= 1$ and satisfies the entropy condition, i.e., $\sum_{i=1}^{k}\hat{p}_i \log \hat{p}_i+\beta =0$. We repeat these steps until the KKT conditions in (\ref{KKT_cond2}) and (\ref{KKT_cond3}) are satisfied.

In Section~\ref{subsect:code_design}, we solve for optimum $\ell(\hat{x}_i)$ for given $\hat{p}_i$; in Section~\ref{subsect:pmf_update}, we solve for optimum $\hat{p}_i$ for given $\ell(\hat{x}_i)$; and in Section~\ref{subsect:C}, we give the iterative algorithm.

\subsection{Age-Optimal Codeword Lengths for a Given PMF} \label{subsect:code_design}
In this section, we find the age-optimal real-valued codeword lengths for a given pmf which satisfy (\ref{KKT_cond2}). First, we show that for a given pmf, the age-optimal codeword lengths should satisfy the Kraft inequality with equality.

\begin{lemma}\label{Lemma:CS}
	For the age-optimal real-valued codeword lengths, we must have $\sum_{i=1}^{k}2^{-\ell(\hat{x}_i)}=1$.
\end{lemma}

\begin{Proof}
	Let us assume, for contradiction, that there exist optimal codeword lengths such that $\sum_{i=1}^{k}2^{-\ell(\hat{x}_i)}<1$. Then, $\theta = 0$ due to (\ref{CS_cond2}). From (\ref{KKT_cond2}), we have
	\begin{align}
	\hat{p}_i\ell(\hat{x}_i)+2\hat{p}_i\mathbb{E}[L] -\lambda p_i = 0, \quad\forall i.
	\end{align}
	By summing over all $i$, we obtain $\mathbb{E}[L] = \frac{\lambda}{3}$. Then, we solve $\ell(\hat{x}_i)= \frac{\lambda}{3}$ for all $i$, which means $p(\lambda) = -\frac{\lambda^2}{9}$. By equating $p(\lambda)$ to zero, we obtain the optimal solution as $\lambda = 0$, which implies $\ell(\hat{x}_i)=0$ for all $i$, which clearly cannot be a solution. Thus, we reach a contradiction, and the age-optimal codeword lengths must satisfy $\sum_{i=1}^{k}2^{-\ell(\hat{x}_i)}=1$.
\end{Proof}

Due to Lemma \ref{Lemma:CS}, we have $\sum_{i=1}^{k}2^{-\ell(\hat{x}_i)}=1$ and $\theta \geq 0$. By summing (\ref{KKT_cond2}) over all $i$, we obtain $\mathbb{E}[L]$ as
\begin{align}\label{exp_length}
\mathbb{E}[L] = \frac{\lambda+\theta \log2}{3}.
\end{align}
We rewrite (\ref{KKT_cond2}), which is,
\begin{align}
-\ell(\hat{x}_i)+ \frac{\theta \log 2}{\hat{p}_i}2^{-\ell(\hat{x}_i)} = 2\mathbb{E}[L]-\lambda
\end{align}
slightly differently as
\begin{align}\label{eqn3}
\frac{\theta (\log2)^2}{\hat{p}_i} 2^{-\ell(\hat{x}_i)} e^{\frac{\theta (\log2)^2}{\hat{p}_i} 2^{-\ell(\hat{x}_i)} }=\frac{\theta (\log2)^2}{\hat{p}_i} 2^{2\mathbb{E}[L]-\lambda}.
\end{align}
Note that (\ref{eqn3}) has the form of $xe^x=y$ for which the solution for $x$ is equal to $x = W(y)$ if $y\geq 0$, where $W(\cdot)$ is the principle branch of the Lambert W function \cite{lambert2}. Using this, we find the unique solution for $\ell(\hat{x}_i)$ as
\begin{align}\label{eqn:opt_lengths}
\ell(\hat{x}_i) = \left( \frac{\lambda-2\theta\log2}{3}\right) +\frac{1}{\log2}W\left(\frac{\theta (\log2)^2}{\hat{p}_i} 2^{\frac{-\lambda +2\theta \log2}{3}} \right).
\end{align}

We note that $\ell(\hat{x}_i)$ in (\ref{eqn:opt_lengths}) has two unknowns in it, $\theta$ and $\lambda$. In order to find the optimal codeword lengths, we find the $(\lambda, \theta)$ pair that satisfies $p(\lambda)  = 0$ and $\sum_{i=1}^{k} 2^{-\ell(\hat{x}_i)} = 1$. Starting from an initial $(\lambda, \theta)$ pair, if $p(\lambda)>0$ (or $p(\lambda)<0$), we increase (or respectively decrease) $\lambda$ in the next iteration as $p(\lambda)$ decreases in $\lambda$. Next, we update $\theta$ by using (\ref{exp_length}). We repeat these steps until $p(\lambda)  = 0$ and $\sum_{i=1}^{k} 2^{-\ell(\hat{x}_i)} = 1$.

\subsection{Age-Optimal PMF for Given Codeword Lengths} \label{subsect:pmf_update}
In this section, we find the age-optimal pmf for given codeword lengths. For this case, we solve $\hat{p}_i$ as
\begin{align}
\hat{p}_i = 2^{\frac{-\frac{1}{2}\ell(\hat{x}_i)^2-2\mathbb{E}[L]\ell(\hat{x}_i)+\lambda \ell(\hat{x}_i)-\sigma}{\gamma}-\frac{1}{\log 2}}. \label{pmf_for_partial}
\end{align}
In order to find the pmf for the partial updates, we solve (\ref{pmf_for_partial}) for a $(\gamma, \sigma)$ pair that satisfies the entropy constraint $ H(\hat{X}) = \beta $ and $\sum_{i=1}^{k} \hat{p}_i = 1$. Starting from an initial $\gamma$, if $H(\hat{X})>\beta$ (or $H(\hat{X})< \beta$), we increase (or respectively decrease) $\gamma$ in the next iteration. Next, we update $\sigma = \frac{\gamma}{\log 2}\left(\log R -1\right)$ where $R = \sum_{i=1}^{k}  2^{\frac{-\frac{1}{2}\ell(\hat{x}_i)^2-2\mathbb{E}[L]\ell(\hat{x}_i)+\lambda \ell(\hat{x}_i)}{\gamma}} $ to ensure $\sum_{i=1}^{k} \hat{p}_i = 1$. We repeat these steps until $ H(\hat{X}) = \beta $ and $\sum_{i=1}^{k} \hat{p}_i = 1$.

\subsection{The Overall Solution} \label{subsect:C}
Using an alternating minimization method \cite{bertsekas, AlterMin, iterminimization2, Niesen07}, starting from an arbitrary pmf, we first find the age-optimal real-valued codeword lengths by following Section~\ref{subsect:code_design}, and then update the pmf by following Section~\ref{subsect:pmf_update}. We repeat this procedure until the first order optimality conditions in (\ref{KKT_cond2}) and (\ref{KKT_cond3}) are satisfied. Since the overall optimization problem in (\ref{problem4_mod}) is not convex, the solution obtained from this iterative alternating minimization algorithm may not be globally optimal.

Finally, recall that, in (\ref{problem4_mod}) and in Section~\ref{subsect:pmf_update}, we solve for the pmf in an unconstrained manner. However, this $k$-point pmf must be such that it can be obtained from the original given $n$-point pmf by combining realizations.
To solve the problem in (\ref{problem3}) optimally, we need to search over all possible partitions of the original updates to generate the pmf for the partial updates, which is NP-hard. Instead, in order to solve this problem practically, especially for large $n$, we apply the proposed alternating minimization technique, which finds a proper pmf and the age-optimal codeword lengths. We then combine updates greedily to find a  partition of the original updates that yields a pmf that approximates the solution obtained by the alternating minimization technique.

\section{Numerical Results} \label{sect:num_res}
\begin{figure}[t]
	\centerline{\includegraphics[width=0.72\columnwidth]{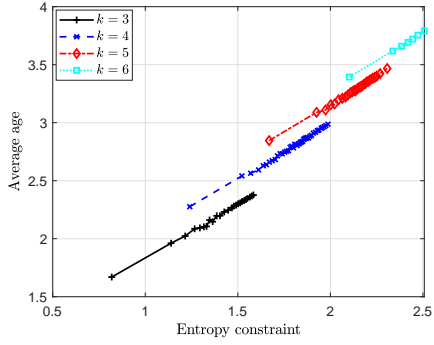}}
	\caption{The optimum average age with real-valued codeword lengths when $X$ is distributed with Zipf$(0.5,8)$ for $k\in \{3,4,5,6\}$.}
	\label{Fig:sim1}
	\vspace{0cm}
\end{figure}

We use Zipf$(s, n)$ as the pmf for the original updates,
\begin{align}
P_X(x_i) = \frac{i^{-s}}{\sum_{j=1}^{n}j^{-s}}, \quad i =1,2,\dots, n.
\end{align}
For the first example, we use Zipf$(0.5,8)$. We vary the entropy constraint $\beta$ and find the corresponding optimum age with real-valued codeword lengths by searching over all possible non-empty partitions of the updates for $k\in \{3,4,5,6\}$. We see in Fig.~\ref{Fig:sim1} that increasing the entropy constraint usually increases the average age. This is similar to classical compression \cite{Cover} where entropy limits the minimum achievable average codeword length, i.e., $H(\hat{X})\leq\mathbb{E}[L]$. Further, we observe in Fig.~\ref{Fig:sim1} that decreasing $k$ achieves a lower average age for a given entropy constraint $\beta$. For example, when the entropy constraint is $\beta = 1.52$, the optimal age is equal to $2.54$ for $k=4$, whereas the optimal age is equal to $2.32$ for $k=3$.

\begin{figure}[t]
	\begin{center}
		\subfigure[\label{c_1}]{%
			\includegraphics[width=0.49\linewidth]{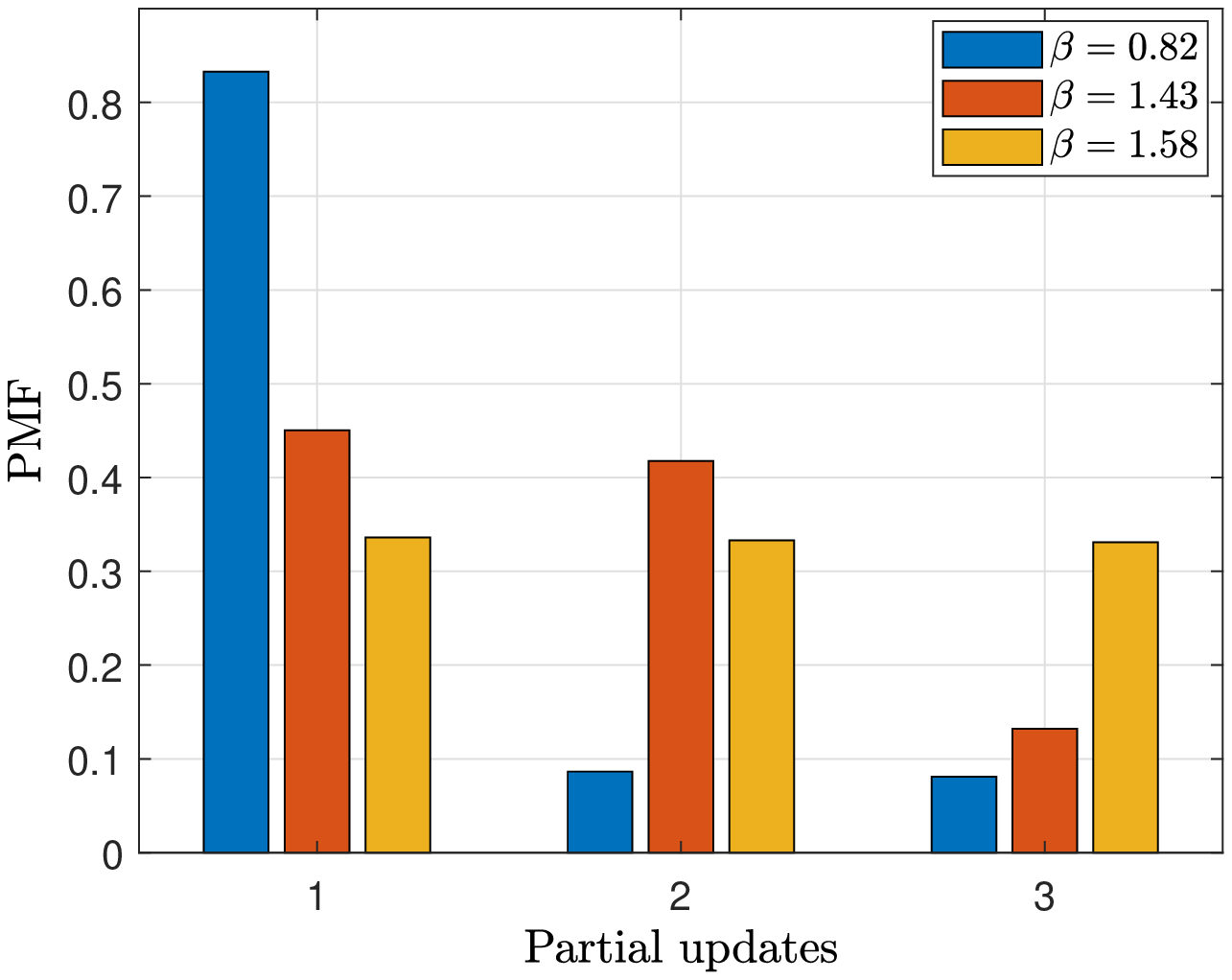}}
		\subfigure[\label{c4}]{%
			\includegraphics[width=0.49\linewidth]{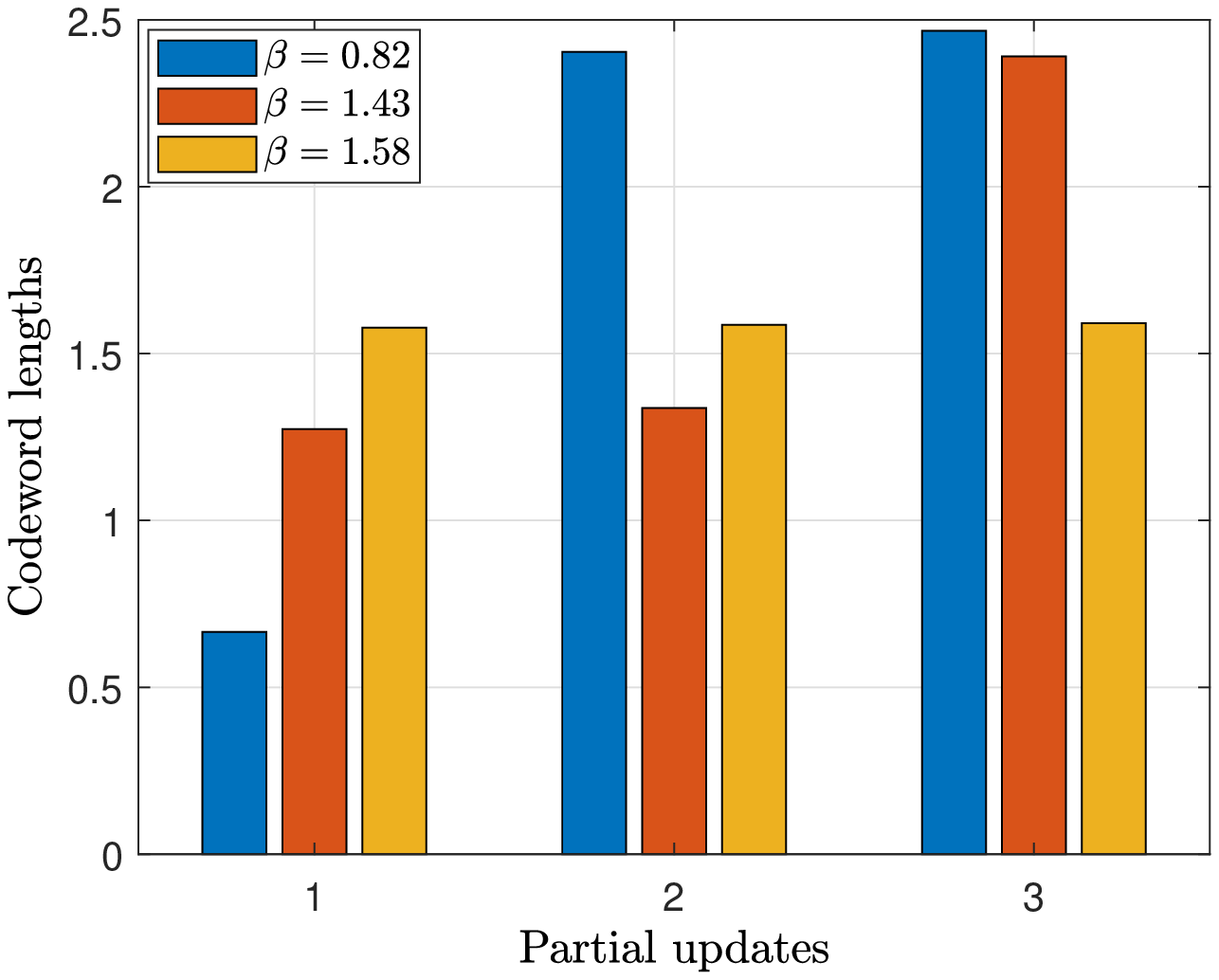}}
		\caption{ We find (a) the age-optimal pmf and (b) the corresponding age-optimal real-valued codeword lengths for $k=3$ with respect to the entropy constraints $\beta\in\{0.82, 1.43, 1.58\}$ for an $X$ with Zipf$(0.5,8)$ distribution.}
		\label{Fig:sim2}
	\end{center}
	\vspace{-0.5cm}
\end{figure}

For the second example, we again use Zipf$(0.5, 8)$ as the pmf for $X$. We find the age-optimal pmf in Fig.~\ref{Fig:sim2}(a) and the corresponding age-optimal real-valued codeword lengths in Fig.~\ref{Fig:sim2}(b) with respect to entropy constraint $\beta\in\{0.82, 1.43, 1.58\}$ when $k=3$. We see that when the entropy constraint is relatively low, i.e., when $\beta=0.82$, the age optimal partial updates are $\hat{\mathcal{X}} =\{\{x_1, x_2,\cdots, x_6\}, x_7, x_8\}$. In other words, the most probable six updates are mapped to one partial update, and the remaining two least probable updates are mapped to the other two partial updates. Since the most probable partial update has the smallest codeword length and realizes most frequently, the system achieves the lowest age compared to other possible partitions. When the entropy constraint is relatively high, i.e., when $\beta= 1.43$ and $\beta = 1.58$, we see in Fig.~\ref{Fig:sim2}(a) that the optimum partial updates are  $\hat{\mathcal{X}} =\{\{x_2,x_4, x_6,x_8\},\{x_1,x_5, x_7\},x_3\}$ and $\hat{\mathcal{X}} =\{ \{x_2,x_6,x_8  \}, \{ x_2,x_3,x_7\}, \{x_1, x_5\} \}$, respectively. Since entropy is a concave function of the pmf, when the entropy constraint is large, the optimum pmf for the partial updates gets closer to a uniform distribution. Thus, age-optimal partition policy strikes a balance between making some partial updates more probable and maintaining the entropy constraint.

\begin{figure}[t]
	\begin{center}
		\subfigure[\label{c_2}]{%
			\includegraphics[width=0.72\linewidth]{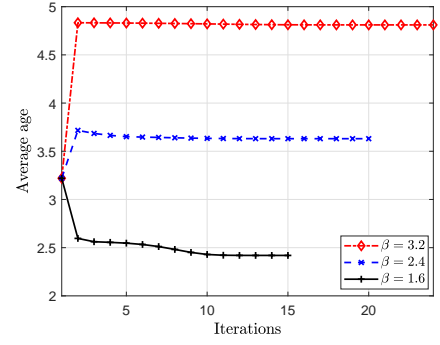}}\\\vspace{-3mm}
		\subfigure[\label{c3}]{%
			\includegraphics[width=0.72\linewidth]{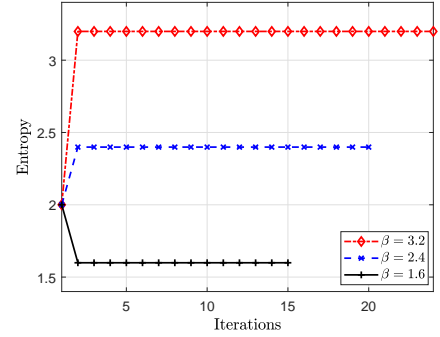}}
		\caption{ We use the proposed alternating minimization method to find the age-optimal pmf and the corresponding age-optimal real-valued codeword lengths for $k=10$ with respect to the entropy constraints $\beta\in\{1.6, 2.4, 3.2\}$ starting from the same arbitrary pmf which has initial entropy close to $2$. We show (a) the age evolution, and (b) entropy evolution, versus iteration index. }
		\label{Fig:sim3}
	\end{center}
	\vspace{-0.5cm}
\end{figure}

In the first two examples, we used exhaustive search over partitions to obtain pmfs $\hat{p}$, and then for each pmf, we found the age-optimal codeword lengths using Section~\ref{subsect:code_design}. For the third example, we use the proposed alternating minimization algorithm to find the pmf and the corresponding age-optimal codeword lengths which satisfy the first order optimality conditions in (\ref{KKT_cond2}) and (\ref{KKT_cond3}) for $k =10$. We use the same initial pmf $P_{\hat{X}}(\hat{x}_i) = \{0.42, 0.32, 0.13, 0.1, 0.02, 0.007, 0.002, 0.0006, 0.00035, \\ 0.00005\}$ for different entropy constraints. When the entropy constraint is relatively low, i.e., when $\beta = 1.6$, we see in Fig.~\ref{Fig:sim3}(a) that the average age initially reduces faster as the entropy reduces to the desired level $1.6$. Then, the average age decreases over the remaining iterations. We find the desired pmf as $P_{\hat{X}}(\hat{x}_i) = \{ 0.3329,0.3329,0.3327,	0.0015,0,\cdots,0 \}$ and its corresponding age-optimal lengths $ \ell(\hat{x}_i)=\{1.59,1.59,1.59,7.55,\infty,\cdots,\infty\}$. We note that even though $k=10$, we see that the pmf has only five partial updates with positive probabilities. This is similar to the result in Fig.~\ref{Fig:sim1}, i.e., decreasing $k$ achieves lower average age with the same $\beta$. Further, the entropy in Fig.~\ref{Fig:sim3}(b) remains the same after iteration two, as the algorithm always forces entropy condition to be satisfied with equality. When $ \beta =2.4$ and $\beta =3.2$, the age increases initially as the entropy increases to the desired level. After the desired entropy is achieved, the average age decreases in the remaining steps. We find $P_{\hat{X}}(\hat{x}_i) = \{ 0.197,0.197,0.197,0.197,0.197,0.015,0,\cdots,0 \}$ and its corresponding age-optimal lengths $ \ell(\hat{x}_i)=\{2.36,2.36,2.36,\\2.36,2.36,5.23,\infty,\cdots,\infty\}$ for $\beta = 2.4$. Further, for $\beta = 3.2$ we find $P_{\hat{X}}(\hat{x}_i) = \{ 0.1107,0.1107,0.1107,0.1107,0.1106,\\0.1106,0.1105,0.1104,0.1101,0.005\}$ and its corresponding age-optimal lengths $ \ell(\hat{x}_i)=\{3.18,3.18,3.18,3.18,3.181,\\3.181,3.182,3.184,3.188,6.857\}$.

\bibliographystyle{unsrt}
\bibliography{IEEEabrv, myLibrary2, lib_v6}

\begin{thebibliography}{10}

\bibitem{Kaul12a}
S.~K. Kaul, R.~D. Yates, and M.~Gruteser.
\newblock Real-time status: How often should one update?
\newblock In {\em IEEE Infocom}, March 2012.

\bibitem{Costa14}
M.~Costa, M.~Codrenau, and A.~Ephremides.
\newblock Age of information with packet management.
\newblock In {\em IEEE ISIT}, June 2014.

\bibitem{Bedewy16}
A.~M. Bedewy, Y.~Sun, and N.~B. Shroff.
\newblock Optimizing data freshness, throughput, and delay in multi-server
  information-update systems.
\newblock In {\em IEEE ISIT}, July 2016.

\bibitem{He16a}
Q.~He, D.~Yuan, and A.~Ephremides.
\newblock Optimizing freshness of information: On minimum age link scheduling
  in wireless systems.
\newblock In {\em IEEE WiOpt}, May 2016.

\bibitem{Sun17a}
Y.~Sun, E.~Uysal-Biyikoglu, R.~D. Yates, C.~E. Koksal, and N.~B. Shroff.
\newblock Update or wait: How to keep your data fresh.
\newblock {\em IEEE Transactions on Information Theory}, 63(11):7492--7508,
  November 2017.

\bibitem{Najm18b}
E.~Najm and E.~Telatar.
\newblock Status updates in a multi-stream {M/G/1/1} preemptive queue.
\newblock In {\em IEEE Infocom}, April 2018.

\bibitem{Najm17}
E.~Najm, R.~D. Yates, and E.~Soljanin.
\newblock Status updates through {M/G/1/1} queues with {HARQ}.
\newblock In {\em IEEE ISIT}, June 2017.

\bibitem{Soysal18}
A.~Soysal and S.~Ulukus.
\newblock Age of information in {G/G/1/1} systems.
\newblock In {\em Asilomar Conference}, November 2019.

\bibitem{Soysal19}
A.~Soysal and S.~Ulukus.
\newblock Age of information in {G/G/1/1} systems: Age expressions, bounds,
  special cases, and optimization.
\newblock May 2019.
\newblock Available on arXiv: 1905.13743.

\bibitem{Yates17b}
R.~D. Yates, P.~Ciblat, A.~Yener, and M.~Wigger.
\newblock Age-optimal constrained cache updating.
\newblock In {\em IEEE ISIT}, June 2017.

\bibitem{Hsu18b}
Y.~Hsu.
\newblock Age of information: Whittle index for scheduling stochastic arrivals.
\newblock In {\em IEEE ISIT}, June 2018.

\bibitem{Kadota18a}
I.~Kadota, A.~Sinha, E.~Uysal-Biyikoglu, R.~Singh, and E.~Modiano.
\newblock Scheduling policies for minimizing age of information in broadcast
  wireless networks.
\newblock {\em IEEE/ACM Transactions on Networking}, 26(6):2637--2650, December
  2018.

\bibitem{Gong19}
J.~Gong, Q.~Kuang, X.~Chen, and X.~Ma.
\newblock Reducing age-of-information for computation-intensive messages via
  packet replacement.
\newblock January 2019.
\newblock Available on arXiv: 1901.04654.

\bibitem{Buyukates19c}
B.~Buyukates and S.~Ulukus.
\newblock Timely distributed computation with stragglers.
\newblock October 2019.
\newblock Available on arXiv: 1910.03564.

\bibitem{Arafa19b}
A.~Arafa, K.~Banawan, K.~G. Seddik, and H.~V. Poor.
\newblock On timely channel coding with hybrid {ARQ}.
\newblock In {\em IEEE Globecom}, December 2019.

\bibitem{Sun17b}
Y.~Sun, Y.~Polyanskiy, and E.~Uysal-Biyikoglu.
\newblock Remote estimation of the {Wiener} process over a channel with random
  delay.
\newblock In {\em IEEE ISIT}, June 2017.

\bibitem{Sun18b}
Y.~Sun and B.~Cyr.
\newblock Information aging through queues: A mutual information perspective.
\newblock In {\em IEEE SPAWC}, June 2018.

\bibitem{Bastopcu19}
M.~Bastopcu and S.~Ulukus.
\newblock Age of information for updates with distortion.
\newblock In {\em IEEE ITW}, August 2019.

\bibitem{bastopcu20}
M.~Bastopcu and S.~Ulukus.
\newblock Age of information for updates with distortion: Constant and
  age-dependent distortion constraints.
\newblock December 2019.
\newblock Available on arXiv:1912.13493.

\bibitem{Zou19b}
P.~Zou, O.~Ozel, and S.~Subramaniam.
\newblock Trading off computation with transmission in status update systems.
\newblock In {\em IEEE PIMRC}, September 2019.

\bibitem{Non_linear}
A.~Kosta, N.~Pappas, A.~Ephremides, and V.~Angelakis.
\newblock Age and value of information: Non-linear age case.
\newblock In {\em IEEE ISIT}, June 2017.

\bibitem{Bastopcu18}
M.~Bastopcu and S.~Ulukus.
\newblock Age of information with soft updates.
\newblock In {\em Allerton Conference}, October 2018.

\bibitem{bastopcu_soft_updates_journal}
M.~Bastopcu and S.~Ulukus.
\newblock Minimizing age of information with soft updates.
\newblock {\em Journal of Communications and Networks}, 21(3):233--243, June
  2019.

\bibitem{Arafa17b}
A.~Arafa and S.~Ulukus.
\newblock Age minimization in energy harvesting communications:
  Energy-controlled delays.
\newblock In {\em Asilomar Conference}, October 2017.

\bibitem{Arafa17a}
A.~Arafa and S.~Ulukus.
\newblock Age-minimal transmission in energy harvesting two-hop networks.
\newblock In {\em IEEE Globecom}, December 2017.

\bibitem{Wu18}
X.~Wu, J.~Yang, and J.~Wu.
\newblock Optimal status update for age of information minimization with an
  energy harvesting source.
\newblock {\em IEEE Transactions on Green Communications and Networking},
  2(1):193--204, March 2018.

\bibitem{Arafa_Age_Online}
A.~Arafa, J.~Yang, and S.~Ulukus.
\newblock Age-minimal online policies for energy harvesting sensors with random
  battery recharges.
\newblock In {\em IEEE ICC}, May 2018.

\bibitem{Arafa18f}
A.~Arafa, J.~Yang, S.~Ulukus, and H.~V. Poor.
\newblock Online timely status updates with erasures for energy harvesting
  sensors.
\newblock In {\em Allerton Conference}, October 2018.

\bibitem{Arafa19e}
A.~Arafa, J.~Yang, S.~Ulukus, and H.~V. Poor.
\newblock Using erasure feedback for online timely updating with an energy
  harvesting sensor.
\newblock In {\em IEEE ISIT}, July 2019.

\bibitem{Farazi18}
S.~Farazi, A.~G. Klein, and D.~R. Brown~III.
\newblock Average age of information for status update systems with an energy
  harvesting server.
\newblock In {\em IEEE Infocom}, April 2018.

\bibitem{Yener_energy_19}
S.~Leng and A.~Yener.
\newblock Age of information minimization for an energy harvesting cognitive
  radio.
\newblock {\em IEEE Transactions on Cognitive Communications and Networking},
  5(2):427--439, May 2019.

\bibitem{Zhong16}
J.~Zhong and R.~D. Yates.
\newblock Timeliness in lossless block coding.
\newblock In {\em IEEE DCC}, March 2016.

\bibitem{Yates_Soljanin_source_coding}
J.~Zhong, R.~D. Yates, and E.~Soljanin.
\newblock Timely lossless source coding for randomly arriving symbols.
\newblock In {\em IEEE ITW}, November 2018.

\bibitem{Mayekar18}
P.~Mayekar, P.~Parag, and H.~Tyagi.
\newblock Optimal lossless source codes for timely updates.
\newblock In {\em IEEE ISIT}, June 2018.

\bibitem{partial_updates}
D.~Ramirez, E.~Erkip, and H.~V. Poor.
\newblock Age of information with finite horizon and partial updates.
\newblock October 2019.
\newblock Available on arXiv:1910.00963.

\bibitem{Zhong17a}
J.~Zhong, E.~Soljanin, and R.~D. Yates.
\newblock Status updates through multicast networks.
\newblock In {\em Allerton Conference}, October 2017.

\bibitem{Zhong18b}
J.~Zhong, R.~D. Yates, and E.~Soljanin.
\newblock Multicast with prioritized delivery: How fresh is your data?
\newblock In {\em IEEE SPAWC}, June 2018.

\bibitem{Buyukates18}
B.~Buyukates, A.~Soysal, and S.~Ulukus.
\newblock Age of information in two-hop multicast networks.
\newblock In {\em Asilomar Conference}, October 2018.

\bibitem{Buyukates18b}
B.~Buyukates, A.~Soysal, and S.~Ulukus.
\newblock Age of information in multihop multicast networks.
\newblock {\em Journal of Communications and Networks}, 21(3):256--267, July
  2019.

\bibitem{Buyukates19}
B.~Buyukates, A.~Soysal, and S.~Ulukus.
\newblock Age of information in multicast networks with multiple update
  streams.
\newblock In {\em Asilomar Conference}, November 2019.

\bibitem{MelihBatu1}
M.~Bastopcu, B.~Buyukates, and S.~Ulukus.
\newblock Optimal selective encoding for timely updates.
\newblock In {\em CISS}, March 2020.

\bibitem{MelihBatu3}
B.~Buyukates, M.~Bastopcu, and S.~Ulukus.
\newblock Optimal selective encoding for timely updates with empty symbol.
\newblock Submitted.

\bibitem{bin_packing}
R.~Kammarti, I.~Ayachi, B.~Pierre, and M.~Ksouri.
\newblock Evolutionary approach for the containers bin-packing problem.
\newblock {\em Studies in Informatics and Control, Informatics and Control
  Publications}, 18(4):315--324, 2009.

\bibitem{frac_programming}
W.~Dinkelbach.
\newblock On nonlinear fractional programming.
\newblock {\em Management Science}, 13(7):435--607, March 1967.

\bibitem{Boyd04}
S.~P. Boyd and L.~Vandenberghe.
\newblock {\em Convex Optimization}.
\newblock Cambridge University Press, 2004.

\bibitem{bertsekas}
D.~P. Bertsekas and J.~Tsitsiklis.
\newblock {\em Parallel and Distributed Computation: Numerical Methods}.
\newblock Englewood Cliffs: Prentice Hall, 1989.

\bibitem{AlterMin}
J.~A. O'Sullivan.
\newblock Alternating minimization algorithms: From {Blahut-Arimoto} to
  expectation-maximization.
\newblock {\em Codes, Curves and Signals:Common Threads in Communications},
  485:173--192, 1998.

\bibitem{iterminimization2}
A.~Yener, R.~D. Yates, and S.~Ulukus.
\newblock Interference management for {CDMA} systems through power control,
  multiuser detection, and beamforming.
\newblock {\em IEEE Transactions on Communications}, 49(7):1227--1239, July
  2001.

\bibitem{Niesen07}
U.~Niesen, D.~Shah, and G.~Wornell.
\newblock Adaptive alternating minimization algorithms.
\newblock In {\em IEEE ISIT}, pages 1641--1645, June 2007.

\bibitem{lambert2}
R.~M. Corless, G.~H. Gonnet, D.~E.~G. Hare, D.~J. Jeffrey, and D.~E. Knuth.
\newblock On the {L}ambert {W} function.
\newblock {\em Advances in Computational Mathematics}, 5(1):329--359, December
  1996.

\bibitem{Cover}
T.~M. Cover and J.~A. Thomas.
\newblock {\em Elements of Information Theory}.
\newblock Wiley Press, 2012.

\end{thebibliography}
\end{document}